\documentclass[twocolumn,aps,prl,superscriptaddress,showpacs]{revtex4}
\usepackage{amsmath,bm}%
\usepackage{graphicx}%

\begin{document}

\draft
%\preprint{}
\title{ Isoscaling of the Fission Fragments with Langevin Equation}

\thanks{ Supported by the Major State Basic Research Development
Program under Contract No G2000774004, the
National Natural Science Foundation of China (NNSFC) under Grant
No 10328259 and 10135030, and the Chinese Academy of Sciences
Grant for the National Distinguished Young Scholars of NNSFC.}

\author{K. Wang }
\affiliation{Shanghai Institute of Applied Physics, Chinese
Academy of Sciences, P. O. Box 800-204, Shanghai 201800}
\affiliation{Graduate School of the Chinese Academy of Sciences}
\author{Y. G. MA} \thanks{Corresponding author. Email: ygma@sinr.ac.cn}
  \affiliation{Shanghai Institute of Applied Physics, Chinese Academy of Sciences, P. O. Box 800-204,
Shanghai 201800}
\author{Y. B. Wei}
\affiliation{Shanghai Institute of Applied Physics, Chinese
Academy of Sciences, P. O. Box 800-204, Shanghai 201800}
\affiliation{Graduate School of the Chinese Academy of Sciences}
\author{X. Z. Cai}
\affiliation{Shanghai Institute of Applied Physics, Chinese
Academy of Sciences, P. O. Box 800-204, Shanghai 201800}
\author{J. G. Chen}
\affiliation{Shanghai Institute of Applied Physics, Chinese
Academy of Sciences, P. O. Box 800-204, Shanghai 201800}
\affiliation{Graduate School of the Chinese Academy of Sciences}
\author{D. Q. Fang}
\affiliation{Shanghai Institute of Applied Physics, Chinese
Academy of Sciences, P. O. Box 800-204, Shanghai 201800}
\author{W. Guo}
\author{G. L. Ma}
\affiliation{Shanghai Institute of Applied Physics, Chinese
Academy of Sciences, P. O. Box 800-204, Shanghai 201800}
\affiliation{Graduate School of the Chinese Academy of Sciences}
\author{W. Q. Shen}
  \affiliation{Shanghai Institute of Applied Physics, Chinese Academy of Sciences, P. O. Box 800-204,
      Shanghai 201800}
 \author{W. D. Tian}
 \affiliation{Shanghai Institute of Applied Physics, Chinese Academy of Sciences, P. O. Box 800-204,
   Shanghai 201800}
\author{C. Zhong}
 \affiliation{Shanghai Institute of Applied Physics, Chinese Academy of Sciences, P. O. Box 800-204,
   Shanghai 201800}
\author{X. F. Zhou}
 \affiliation{Shanghai Institute of Applied Physics, Chinese Academy of Sciences, P. O. Box 800-204,
   Shanghai 201800}
\affiliation{Graduate School of the Chinese Academy of Sciences}

\date{\today}

\begin{abstract}
Langevin equation is used to simulate the fission process of
$^{112}$Sn + $^{112}$Sn and $^{116}$Sn + $^{116}$Sn. The mass
distribution of the fission fragments  are given by assuming the
process of symmetric fission or asymmetric fission with the
Gaussian probability sampling. Isoscaling behavior has been
observed from the analysis of fission fragments of both reactions
and the isoscaling parameter $\alpha$ seems to be sensitive to the
width of fission probability and the beam energy.
\end{abstract}

\pacs{24.75.+i, 25.85.Ge, 21.10.Tg}

\keywords{Isoscaling, Langevin equation, isotope distribution,
nuclear fission }

\maketitle

%\textbf{I.Introduction}

Since the isoscaling law has been observed experimentally
\cite{TsangPRL, Tsang2,Tsang3}, many statistical models have
successfully explained the isoscaling behavior. Isoscaling means
that the ratio of isotope yields from two different reactions, 1
and 2, $R_{21}(N,Z)=Y_2(N,Z)/Y_1(N,Z)$, is found to exhibit an
exponential relationship as a function of the neutron number $N$
and proton number $Z$ \cite{TsangPRL}
\begin{equation}
 R_{21}(N,Z) = \frac{Y_2(N,Z)}{Y_1(N,Z)} = C exp(\alpha N + \beta Z).
\end{equation}
where  $C$, $\alpha$ and $\beta$ are three parameters. In
grand-canonical limit, $\alpha = \Delta\mu_{n}/T$ and $\beta =
\Delta\mu_{z}/T$ where $\Delta\mu_{n}$ and $\Delta\mu_{z}$ are the
differences between the neutron and proton chemical potentials for
the two reactions, respectively. This behavior is attributed to
the difference of two reaction systems with different isospin
asymmetry. It is potential to probe the isospin dependent nuclear
equation of state by the studies of isoscaling \cite{Ma_review}.
So far, the isoscaling behavior has been  experimentally explored
by various reaction mechanisms, ranging from the evaporation
\cite{TsangPRL}, fission \cite{Friedman,Veselsky2} and deep
inelastic reaction at low energies to the projectile fragmentation
\cite{TAMU1,Veselsky} and multi-fragmentation at intermediate
energy \cite{TsangPRL,LiuTX,Geraci}. While, the isoscaling
phenomenon has been extensively examined in different theoretical
frameworks, ranging from dynamical model, such as BUU model
\cite{LiuTX} and anti-symmetrical molecular dynamics model
\cite{Ono}, to statistical models, such as the expansion emission
source model, statistical multi-fragmentation model and the
lattice gas model \cite{Tsang2,Tsang3,Botvina,Souza,Ma}.

In this work, we present an analyse for the fragments from the
fission which was simulated by Langevin equation. The isotopic or
isotonic ratios of the different fragment yields from $^{116}$Sn +
$^{116}$Sn and $^{112}$Sn + $^{112}$Sn system are presented and
the features of the isoscaling behavior in fission dynamics are
investigated.

The process of fission can be described in terms of collective
motion using the transport theory
\cite{Randrup,Feldmeier,HHofmann,Frobrich}. The dynamics of the
collective degrees of freedom is typically described using the
Langevin or Fokker-Planck equation. In this Letter,  we deal with
a Combine Dynamical and Statistical Model (CDSM) which is a
combination of a dynamical Langevin equation and a statistical
model to describe the fission process of heavy ion reaction. This
model is an overdamped Langevin equation coupled with a Monte
Carlo procedure allowing for the discrete emission of light
particles. It switches over to statistical model when the
dynamical description reaches a quasi-stationary regime. We first
specify the entrance channel through which a compound nucleus is
formed, ie. the target and projectile is complete fusion.

In this work the total initial excitation energy $E^{*}_{tot}$ is
given by $E^{*}_{tot} = E_{lab}A_{T}/(A_{T}+A_{P})+ Q$ where
$A_{T}$ and $A_{P}$ represents the mass of target and projectile,
respectively, and  $Q$ is the fusion $Q$-value calculated by $Q =
M_{T}+M_{P}-M^{LD}_{CN}$. $M_{T}$ and $M_{P}$ is the mass of
projectile and target come from experimental data, respectively.
If it is unavailable, it is calculated by macroscopic-microscopic
model \cite{macr-micr}. $M^{LD}_{CN}$ is the mass of the compound
nucleus which is calculated from the liquid-drop model.

The dynamical part of CDSM model is described by Langevin equation
which is driven  by the free energy $F$. $F$ is related to the
level density parameter $a(q)$ \cite{Ignatyuk}
\begin{equation}
F(q,T) = V(q) - a(q)T^{2}
\end{equation}
in the Fermi gas model, where $V(q)$ is fission potential.

The overdamped  Langevin equation reads
\begin{eqnarray}
 \frac{dq}{dt} = -\frac{1}{M\beta(q)}(\frac{\partial F(q,T)_{T}}{\partial
 q})+\sqrt{D(q)}\Gamma(t),
\end{eqnarray}
where $q$ is the dimensionless fission coordinate defined as half
of the distance between the centers of masses of the future
fission fragments. $\Gamma(t)$ is a time-dependent stochastic
variable with Gaussian distribution. Its average and  correlation
function is written as
\begin{eqnarray}
 <\Gamma(t)> = 0,\nonumber\\
 <\Gamma(t)\Gamma(t')> = 2\delta_{\varepsilon}(t-t').
\end{eqnarray}
The fluctuation strength coefficient $D(q)$ can be expressed
according to the fluctuation-dissipation theorem:
\begin{equation}
D(q) = \frac{T}{M\beta_0(q)} ,
\end{equation}
where $M$ is the total mass and $\beta_0(q)$ is the reduced
friction parameter which is the only parameter of this model.

The potential energy $V(A,Z,L,q)$ is obtained from the
finite-range liquid drop model \cite{liquid}
\begin{eqnarray}
V(A,Z,L,q) = a_{2}[1-k(\frac{N-Z}{A})^{2}]A^{2/3}[B_{s}(q)-1]\nonumber\\
 +
 c_{3}\frac{Z^{2}}{A^{1/3}}[B_{c}(q)-1]+c_{r}L^{2}A^{-5/3}B_{r}(q),
\end{eqnarray}
where $B_{s}(q)$, $B_{c}(q)$ and $B_{r}(q)$ means surface, Coulomb
and rotational energy terms, respectively, which depends on the
deformation coordinate $q$. $a_{2}$, $c_{3}$, $k$ and $c_{r}$ are
parameters not related to $q$. In our calculation we take them
according to Ref.~\cite{Frobrich}.

We use $c$ and $h$ \cite{ch} to describe the shape of nucleus,
\begin{equation}
\rho^{2}(z) =
(1-\frac{z^{2}}{c_{0}^{2}})((\frac{1}{c^{3}}-\frac{b_{0}}{5})c_{0}^{2}+
B_{sh}(c,h)z^{2}),
\end{equation}
where
\begin{equation}
{c_{0}} = c R, ~~~~R = 1.16A^{1/3}.
\end{equation}
The nuclear shape function $B_{sh}(c,h)$ and the collective
fission coordinate $q(c,h)$ of mass number $A$ is expressed as
\begin{eqnarray}
 B_{sh}(c,h) = 2h + \frac{c-1}{2},\nonumber\\
 q(c,h) = \frac{3}{8}c(1 + \frac{2}{15}B_{sh}(c,h)c^{3}).
\end{eqnarray}

The fission process of Langevin equation is propagated using an
interpretation of Smoluchowski \cite{interpretation}. In our
calculation we adopt one-body dissipation (OBD) friction form
factor $\beta_0(q)$ \cite{obd} which is calculated with one-body
dissipation with a reduction of wall term. Here we use an
analytical fit formula which was developed in Ref. \cite{fit}
\[\beta_{OBD}(q)=\left.\{ \begin{array}{ll}15/q^{0.43} + 1 -
10.5q^{0.9} + q^{2} & \mbox{if $q>0.38$}\\32-32.21q & \mbox{if
$q<0.38$}\end {array} \right. \]

In the dynamical part of the model the emission of light particles
($n,\alpha, p, d$) and giant dipole $\gamma$ are calculated at
each Langevin time step $\tau$, the widths for particle and giant
dipole $\gamma$ decay are given by the parametrization of Blann
\cite{Width} and Lynn \cite{Lynn}, respectively.

Within the framework of  Langevin simulation we chose 200,000
fission events which happen on dynamic channel (we give up the
events which happen in statistic part of CDSM model) and chose a
Gaussian distribution random number as the mass asymmetry
parameter $\alpha_0 = (A_{1}-A_{2})/(A_{1}+A_{2})$, which is
defined as the ratio of the volumes of two parts of the nucleus
obtained when it reaches the scission point. When $\alpha_0 = 0$
means symmetrical fission. It is taken from a Gaussian
distribution random number from -1 to 1 with the mean is 0.
$A_{1}$ and $A_{2}$ is the mass of the two fission fragments,
respectively. In this work we assume the fission fragments have
the same $N/Z$ as the initial system and then $Z_1$ or $Z_2$ of
fission fragments can be deduced from $A_{1}$ or $A_2$. This
assumption is similar to the case of of deep inelastic heavy ion
collisions at low energies, where the isospin degree of freedom
has been found to first reach equilibrium \cite{isospin}.

From a practical point of view, the isoscaling occurs when  two
mass distributions for a given $Z$ from two processes with
different isospin are  Gaussian distributions with the same width
but different mean mass. Basically, the isotopic distribution can
be described by
\begin{equation}
Y(N,Z) = f(Z) exp[-\frac{(N-N_Z)^2}{2\sigma_Z^2}],
\end{equation}
where $N_Z$ is the centroid of the distribution and $\sigma_Z^2$
describes the variance of the distribution for each element of
charge $Z$. This leads to an exponential behavior of the ratio
$R_{21}$ if the quadratic term in $N_Z$ is neglected,
\begin{equation}
ln(R_{21}) \sim  \frac{[(N_Z)_2-(N_Z)_1] N}{\sigma_Z^2}.
\label{appro}
\end{equation}
Note that Eq.(\ref{appro}) requires the values for the
$\sigma_Z^2$ to be approximately the same for both reactions,
which is a necessary condition for isoscaling. Indeed, we observed
this case in our simulations for Sn + Sn collisions. In the
Langevin equation, $\sigma_Z^2$ mainly depends  on the physical
conditions reached, such as the temperature, the density and the
friction parameter etc. Considering that $R_{21}' = exp(\alpha N)$
for a given $Z$, $\alpha \sim
\frac{[(N_Z)_2-(N_Z)_1]}{\sigma_Z^2}$. Usually $\sigma_Z^2$ can be
considered to be proportional to temperature $T$ of the fragments,
in this way
\begin{equation}
\alpha \sim \frac{[(N_Z)_2-(N_Z)_1]}{T},
\end{equation}
where $[(N_Z)_2-(N_Z)_1]$ can be understood as a term of the
chemical potential difference between two reactions.

\begin{figure}
\vspace{-0.9truein}
\includegraphics[scale=0.40]{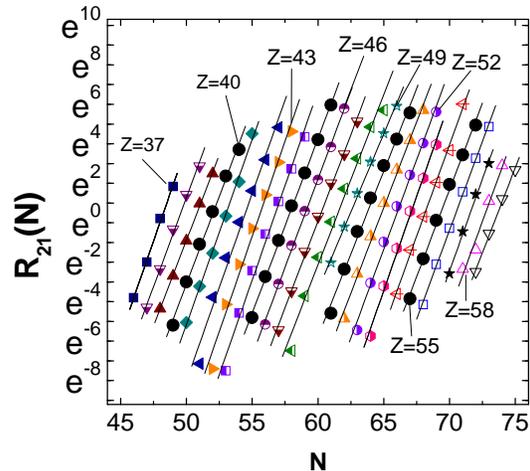}
\vspace{-0.6truein}
 \caption{\footnotesize The yield ratio of the fission fragments
 between $^{116}Sn+^{116}Sn$ and  $^{112}Sn+^{112}Sn$ in the
Langevin model with $\sigma_{\alpha_0} = 0.06$ and  E/A = 8.4 MeV.
Different symbols from left to right represent the calculated
results for the isotopes from Z = 37 to 59.  The lines represent
exponential fits to guide the eye.} \label{fig_isoscaling}
\end{figure}

\begin{figure}
\vspace{-1.0truein}
\includegraphics[scale=0.35]{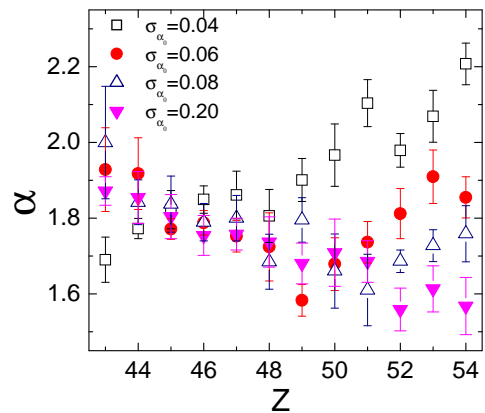}
\vspace{-0.6truein}
 \caption{\footnotesize The scaling parameter $\alpha$ as
a function of $Z$ in the different width ($\sigma_{\alpha_0}$) of
the mass asymmetry parameter $\alpha_0$ with Gaussian distribution
for fission. } \label{fig_alpha_N_sgmX}
\end{figure}

Eq.(1) can be written as $lnR_{21} = C_Z + \alpha N$, where $C_Z =
lnC + \beta Z$, if we plot $R_{21}$ as function of N, on a natural
logarithmic plot, the ratio follows  along a straight line. In
Fig.~\ref{fig_isoscaling} this behavior is observed in Langevin
simulation. From the figure, the relationship between  $\alpha$
and the charge number $Z$ of the fission fragments can be deduced.
In order to investigate the effect of the width of Gaussian
probability distribution on the isoscaling behavior, we change the
widths of the Gaussian distribution of mass asymmetry parameter
$\alpha_{0}$, namely $\sigma_{\alpha_0}$ = 0.04, 0.06, 0.08 and
0.20, with the random number from -1 to 1 and the mean of 0.
Fig.\ref{fig_alpha_N_sgmX} shows the isoscaling parameter $\alpha$
as a function of $Z$ in the conditions of different
$\sigma_{\alpha_0}$. From this figure, we know in the low
$\sigma_{\alpha_0}$, i.e., the fission fragments are
overwhelmingly dominated by the symmetric fission, $\alpha$
increases with $Z$. This means that the isospin effect becomes
stronger with the increasing of $Z$. In a recent analyse of
Friedman \cite{Friedman} with a simple liquid-drop model, a
systematic increase of the isoscaling parameter $\alpha$ with the
proton number of the fragment element is predicted.  In our
simulation, this kind of increase of $\alpha$ with $Z$ stems from
the dominated symmetric fission mechanism. While, in the another
extreme case from the Fig.\ref{fig_alpha_N_sgmX}, i.e. with the
larger $\sigma_{\alpha_0}$, $\alpha$ shows a contrary trend with
$Z$, i.e., it drops with $Z$. In this case, it seems that there
exists stronger isospin effect for the fragments with lower $Z$.
In a middle case, the rising branch and falling branch competes
with each other, the mediate isoscaling behavior appears and a
minimum of $\alpha$ parameter occurs around the symmetric fission
point. We note that the fission data of $^{238,233}$U targets
induced by 14 MeV neutrons reveals the backbending behavior of the
isoscaling parameter $\alpha$ around the symmetric fission point
\cite{Veselsky2} as stated above. In their study, they interpreted
this originates from the temperature difference of fission
fragments since the isoscaling parameter is typically, within the
grand-canonical approximation, considered inversely proportional
to the temperature ( $\alpha$ = $\Delta \mu_{n}$/T ) as stated
above. It is, however, not $a ~priori$ obvious why such  a
grand-canonical formula can be applied to fission. In our case,
this kind of backbending of isoscaling parameter apparently stems
from the moderate width of the fission probability of the
fissioning nucleus as shown in Fig.\ref{fig_alpha_N_sgmX}. Of
course, we did not exclude the change of temperature of the
fission fragments due to the change of the width of the fission
probability. Overall speaking, we find that the isoscaling
behavior  is sensitive to the width of the fission probability
distribution.

\begin{figure}
\vspace{-0.7truein}
\includegraphics[scale=0.35]{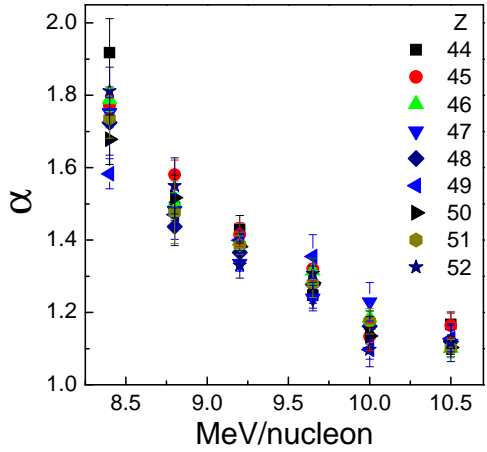}
\vspace{-.5truein} \caption{\footnotesize $\alpha$ as a function
of beam energy for the fragments Z = 44-52. The width of the
Gaussian probability $\sigma_{\alpha_0}$ is  0.06.  } \label{fig3}
\end{figure}

In addition, the simulations are systematically done in different
beam energies. The values of $\alpha$ are extracted as a function
of beam energy for the fragments Z = 44-52 as shown in
Fig.\ref{fig3}. It shows that $\alpha$ decreases as the beam
energy increasing which means that the isospin effect fades away
with the increasing of $E_{lab}$. This behavior is similar to that
 $\alpha$ drops with the temperature in the statistical models as
well as experiments \cite{Tsang3,Ma,Ma_1999,Ma_CP}.

In summary, we applied Langevin  model to investigate the
isoscaling behavior in the dynamical process of compound nuclear
fission. In order to treat the fission fragments, we assume that
the mass asymmetry parameter of the fissioning nucleus is taken
from a random number of Gaussian distribution whose width is
$\sigma_{\alpha_0}$. The simulation illustrates that the yield
ratios of fission fragments in the dynamical fission of $^{116}Sn
+ ^{116}Sn$ and $^{112}Sn+^{112}Sn$ reaction system shows the
isoscaling behavior. More interestingly, we found that the
isoscaling parameter $\alpha$ is sensitive strongly to the
Gaussian width $\sigma_{\alpha_0}$ of the mass asymmetry
parameter. When $\sigma_{\alpha_0}$ is small, i.e. the fission is
almost symmetrical, $\alpha$ increases with the atomic number of
fission fragments, which is similar to the theoretical prediction
of a simple liquid-drop model \cite{Friedman}. In contrary, when
$\sigma_{\alpha_0}$ is large, for instance, $\sigma_{\alpha_0}$ =
0.20, $\alpha$ drops with $Z$ of fission fragments. However, in
the intermediate values of $\sigma_{\alpha_0}$, $\alpha$ shows a
backbending with $Z$ of fission fragments, which is similar to the
observation of the $^{238,233}$U fission data induced by 14 MeV
neutrons \cite{Veselsky2}. In addition, it is found that $\alpha$
drops with the beam energy of the projectile, reflecting the
temperature dependence of isoscaling parameter. In general, the
isoscaling analysis of the fission data appears to be a sensitive
tool to investigate the fission dynamics.

{}


\footnotesize
\begin{thebibliography}{}
\bibitem{TsangPRL}  Tsang M B et al. 2001 {\it Phys. Rev. Lett.} {\bf 86} 5023.
\bibitem{Tsang2} Tsang M B et al. 2001 {\it Phys. Rev.}  C {\bf 64} 041603.
\bibitem{Tsang3} Tsang M B  et al. 2001 {\it Phys. Rev.} C {\bf 64} 054615.
\bibitem{Ma_review}Ma Y G et al. 2002 {\it Prog. Phys.} (in Chinese)
{\bf 22} 99; \\ Ma Y G et al. 2004 {\it Nucl. Sci. Tech.} {\bf 15}
4.

\bibitem{Friedman}Friedman W A  2004 {\it Phys. Rev.} C {\bf 69}
031601(R).

\bibitem{Veselsky2} Veselsky M, Souliotis G A, and Jandel M 2004
{\it Phys. Rev.} C {\bf 69} 044607.

\bibitem{TAMU1}Souliotis G A et al. 2003 {\it Phys. Rev.} C {\bf 68} 024605.
\bibitem{Veselsky} Veselsky M et al. 2004 {\it Phys. Rev.} C {\bf 69} 031602(R).
\bibitem{LiuTX}Liu T X et al. 2004 {\it Phys. Rev.} C {\bf 69} 014603.
\bibitem{Geraci}Geraci E et al. 2004 {\it  Nucl. Phys.} A {\bf 732} 173.

\bibitem{Ono}Ono A et al. 2003 {\it  Phys. Rev.} C {\bf 68} 051601.

\bibitem{Ma} Ma Y G et al. 2004 {\it Phys. Rev.} C {\bf 69} 064610.

\bibitem{Botvina}Botvina A S, Lozhkin O V, and  Trautmann W 2002 {\it Phys. Rev.} C {\bf 65} 044610.

\bibitem{Souza}Souza  S  R et al. 2004 {\it  Phys. Rev.} C {\bf 69} 031607(R).
\bibitem{Frobrich}Fr\"obrich P, Gontchar I I 1998 {\it  Phys. Rep.} {\bf 292} 131.
\bibitem{Randrup}Randrup J 1979 {\it Nucl. Phys.} A {\bf 327} 490.

\bibitem{Feldmeier}Feldmeier H 1987 {\it Rep. Prog. Phys.} {\bf 50} 915.

\bibitem{HHofmann}Hofmann H 1997 {\it Phys. Rep.} {\bf 284} 137.

\bibitem{macr-micr} M\"{o}ller P et al. 1998 {\it At. Data \& Nucl.
Data Tables} 39 225.

\bibitem{Ignatyuk}Ignatyuk A V et al. 1985 {\it Fiz. EJ. Chast. At.
Yadra} {\bf 16} 709.

\bibitem{liquid}Myers W D, Swiatecki W J 1996
{\it Nucl. Phys} {\bf 81} 1;\\ Myers W D, Swiatecki W J 1967 {\it
Ark Fys.} {\bf 36} 343.

\bibitem{ch}Hasse R W,  Myers W D 1988 {\it  Geometrical Relationships of
Macroscopic Nuclear Physics} (Spinger, Berlin, Heidelberg, New
York).

\bibitem{interpretation}Klimontovich Yu L 1994 Nonlinear Brownian
motion, Physics Uspekhi {\bf 37}, 737.

\bibitem{obd}Blocki J et al. 1978 {\it Ann. Phys.} {\bf 113} 330.

\bibitem{fit}Gontchar I and  Litnevsky L A 1997 {\it Z. Phys.  } A {\bf 26} 347.

\bibitem{Width}Blann M 1980 {\it  Phys. Rev.} C {\bf 21} 1770.

\bibitem{Lynn}Lynn J E 1968 {\it Theory of Neutron Resonance
Reactions} (Clarendon, Oxford).
\bibitem{isospin} Gatty B et al. 1975 {\it Z. Phys.} A {\bf 273} 65.

\bibitem{Ma_1999} Ma  Y G et al. 1999 {\it  Chin. Phys. Lett.} {\bf 16}
256;\\ Ma Y G et al. 1999 {\it Phys. Rev.} C {\bf 60} 024607.
\bibitem{Ma_CP}Ma Y G 1999 {\it Acta Phys. Sin.} {\bf 49 } 654.


\end{thebibliography}
\end{document}